\documentclass{bmcart}

\RequirePackage{hyperref}
\usepackage[utf8]{inputenc} 

\usepackage{amsthm,amsmath}
\usepackage{todonotes}
\usepackage{mathtools}

\DeclarePairedDelimiter\floor{\lfloor}{\rfloor}
\usepackage[capitalise]{cleveref}
\usepackage{algorithm2e}
\usepackage{siunitx}
\crefname{algocf}{Algorithm}{Algorithms}
\crefname{thm}{Theorem}{Theorem}
\usepackage{gensymb}
\usepackage{amsfonts}
\usepackage{float}
\newcounter{thm}
\newtheorem{Theorem}[thm]{Theorem}
\usepackage{filecontents}

\startlocaldefs
\endlocaldefs

\begin{document}

\begin{frontmatter}

\begin{fmbox}
\dochead{Research}


\title{Isolating phyllotactic patterns embedded in the secondary growth of sweet cherry ({\it Prunus avium} L.) using magnetic resonance imaging}
\author[
   addressref={cmse},
   email={eithunmi@msu.edu}
]{\inits{ME}\fnm{Mitchell} \snm{Eithun}}
\author[
   addressref={hort},
   email={larso102@mail.msu.edu}
]{\inits{JL}\fnm{James} \snm{Larson}}
\author[
   addressref={hort},
   email={langg@msu.edu}
]{\inits{GL}\fnm{Gregory} \snm{Lang}}
\author[
   addressref={hort,cmse},
   email={dhchitwood@gmail.com}
]{\inits{DHC}\fnm{Daniel H.} \snm{Chitwood}}
\author[
   addressref={cmse,math},
   email={muncheli@egr.msu.edu}
]{\inits{EM}\fnm{Elizabeth} \snm{Munch}}


\address[id=cmse]{
  \orgname{Department of Computational Mathematics, Science and Engineering, Michigan State University}, 
 \city{East Lansing, MI},
  \cny{USA}
  \postcode{48823}
}
\address[id=hort]{
  \orgname{Department of Computational Mathematics, Science and Engineering, Michigan State University}, 
 \city{East Lansing, MI},
  \cny{USA}
  \postcode{48823}
}
\address[id=math]{
  \orgname{Department of Mathematics, Michigan State University}, 
 \city{East Lansing, MI},
  \cny{USA}
  \postcode{48823}
}


\begin{artnotes}
\end{artnotes}

\end{fmbox}


\begin{abstractbox}

\begin{abstract} 

\textbf{Background}: Epicormic branches arise from dormant buds patterned during the growth of previous years. Dormant epicormic buds remain on the surface of trees, pushed outward from the pith during secondary growth, but maintaining vascular connections. Epicormic buds can be reactivated, either through natural processes or intentionally, to rejuvenate orchards and control tree architecture. Because epicormic structures are embedded within secondary growth, tomographic approaches are a useful method to study them and understand their development.

\textbf{Results}: We apply techniques from image processing to determine the locations of epicormic vascular traces embedded within secondary growth of sweet cherry (\textit{Prunus avium} L.), revealing the juvenile phyllotactic pattern in the trunk of an adult tree. Techniques include breadth-first search to find the pith of the tree, edge detection to approximate the radius, and a conversion to polar coordinates to threshold and segment phyllotactic features. Intensity values from Magnetic Resonance Imaging (MRI) of the trunk are projected onto the surface of a perfect cylinder to find the locations of traces in the ``boundary image''. Mathematical phyllotaxy provides a means to capture the patterns in the boundary image by modeling phyllotactic parameters. Our cherry tree specimen has the conspicuous parastichy pair $(2,3)$, phyllotactic fraction 2/5, and divergence angle of approximately \ang{143} degrees. 

\textbf{Conclusions}: The methods described not only provide a framework to study phyllotaxy, but for image processing of volumetric image data in plants. Our results have practical implications for orchard rejuvenation and directed approaches to influence tree architecture. The study of epicormic structures, which are hidden within secondary growth, using tomographic methods also opens the possibility of studying the genetic and environmental basis of such structures. 

\end{abstract}




\end{abstractbox}
%

\end{frontmatter}



\section{Introduction}
Plants increase in length from apical meristems during primary growth. Located at the shoot tip, the shoot apical meristem is the site of cell division that produces leaf and bud primordia. Newly divided cells elongate, pushing the apical meristem upward. As the shoot continues to grow, leaf primordia cells divide, differentiate, and expand into leaves that subtend axillary buds. The point at which each leaf is attached to the shoot constitutes a node. In sweet cherry (\textit{Prunus avium} L.), leaves form at each node in the year that a shoot develops. In non-juvenile plants, the axillary buds at nodes at the base of the shoot may begin differentiating into solitary flower buds. These bloom, develop fruit if pollinated and fertilized, and upon abscission of the fruit, the node becomes void of apparent vegetative or reproductive buds for subsequent growth. These are called ``blind'' nodes. The remaining non-basal, majority of nodes on the new shoot form a single vegetative bud at each node. In spring of the year after the shoot's formation, each of these buds usually produces five to eight leaves that arise from very closely-packed nodes, producing a spur (a short, modified branch) along the rest of the length of the shoot, except for the terminal shoot apical meristem that again elongates to form a new section of shoot. Buds on some of the uppermost nodes of the original shoot also may elongate into new lateral shoots, rather than form spurs \cite{quero-garcia_cherries:_2017}. In the orchard, trees are considered mature once they have filled their allotted orchard space and have all of their reproductive components; in modern, high density plantings, this is three to five years.

The shoot apical meristem forms nodes in spiral patterns resembling cylindrical helices. \textit{Phyllotaxy} is the study of the arrangement of plant organs during their development. These organs can be branches, leaves, vascular traces, or other features associated with nodes. Frequently, organ primordia form spiral patterns called \textit{parastichies}, which can be characterized by Fibonacci-type sequences of numbers.  For centuries, mathematicians have been interested in building models to describe the geometry of phyllotactic patterns \cite{jean_phyllotaxis_1994}. In these models, \textit{phyllotactic parameters} are numbers that capture the layout and spacing of primordia characteristic of different plant species and their stages of development \cite{i._vakarelov_method_2018}. 

Phyllotactic patterns are often studied in two forms: centric and cylindrical \cite{jean_phyllotaxis_1994}. In the \textit{centric} representation, parastichy spirals emanate from a central point, like the capitulum of a sunflower.  In the \textit{cylindrical} model, parastichies are helices on the surface of a cylinder, like in pineapples or pinecones. In the cylindrical representation, it is convenient to unwrap the surface of the cylinder and view the primordia as a cylindrical lattice in the plane, called a \textit{Bravais lattice}, in which intersections of parastichies are primordia.

After a shoot has elongated and the phyllotaxy of the nodes patterned, woody plants undergo secondary growth to increase in girth. Axillary buds usually remain dormant in the year of initiation, but when the primary shoot is damaged or is growing extremely rapidly, the axillary bud may elongate into a new lateral shoot. Axillary buds that remain dormant will eventually become engulfed by secondary growth of the stem and persist beneath the bark as an epicormic bud meristem \cite{busgen_structure_1929,stone_dormant_1943}. Epicormic bud meristem cells divide and elongate with radial growth, maintaining their presence just beneath the bark, and leaving a vascular trace back to the pith, shown in \cref{fig:parts}. Epicormic traces are 2-5 mm high \cite{fontaine_ontogeny_1998} and occur perpendicular to the pith \cite{busgen_structure_1929,colin_tracking_2010}. Epicormic buds may sprout into epicormic branches following a stress such as fire \cite{burrows_syncarpia_2008}, insect defoliation \cite{piene_spruce_1996}, wind damage \cite{cooper-ellis_forest_1999}, competition \cite{nicolini_stem_2001}, or pruning \cite{ohara_epicormic_2009}. If the primary epicormic meristem becomes damaged, it may split via the initiation of subtending secondary bud meristems. Epicormic buds can be reactivated under the right conditions, and serve as a reserve of future potential shoot growth that can be used to rejuvenate orchards by farmers.
 
Image processing techniques have previously been used for dendochronology, the study of tree rings and their features \cite{levanic_atrics_2009,sundari_approach_2017} and to manually locate rameal traces in oak trunks \cite{colin_tracking_2010}. Semiautomatic feature identification on trees using image processing has been used for tree ring identification using the Sobel operator \cite{henke_semiautomatic_2014}. In this work, we isolate the patterning of epicormic traces embedded in secondary growth from a magnetic resonance imaging (MRI) scan of an eight year old sweet cherry tree.
An image processing framework is implemented that finds the pith and radius of the trunk across slices. Using these dimensions, a polar coordinate conversion of each slice reveals x-ray dense regions corresponding to epicormic traces. A blob detection algorithm segments the epicormic features and the resulting phyllotactic parameters are estimated. Our work reveals the juvenile phyllotactic pattern of a sweet cherry tree embedded within 8 years of secondary growth. The study of epicormic features has implications for orchard rejuvenation, and the analysis methods presented provide an empirically based method to measure phyllotactic patterning and isolate anatomical features in plants.


\section{Materials and Methods}
\label{sec:Bkgd}

\subsection{Plant material and Magnetic Resonance Imaging (MRI)}
At bloom (late April) in 2016, the top of an 8-year-old sweet cherry tree, cultivar 'NY 119', planted at Michigan State University's Clarksville Research Center and trained to a standard central leader canopy architecture, was removed with a chain saw 1.5 m from the ground. All lateral branches below that point were removed to promote the sprouting of epicormic buds along the remaining trunk length. Prior to and following this topping, the tree was managed with standard fertility, irrigation, and pest management procedures with the rest of the $\sim 0.25$ ha experimental orchard. In August, a 1.14 m-long section of trunk was fully removed just above the graft union and dried at room temperature in the laboratory.  In December, this trunk section was scanned at slice thickness set of 0.625 mm, at Michigan State University’s Department of Radiology (East Lansing, MI) using a whole body magnetic resonance scanner (GE Signa HDX 3.0T, Chicago, IL). The scan took slightly over a minute and produced 1871 images, each of which is $250 \times 250$ pixels. Contrast in the scanned image is created by differences in moisture content; contrast is greater in air-dried than fresh wood \cite{freyburger_measuring_2009}. The specimen and the scan are shown in \cref{fig:tree}. \cref{tab:dimensions} gives estimates for the size and shape of the specimen.



\begin{table}[h!]
\centering
\begin{tabular}{ll}
\textbf{Measurement} & \textbf{Value} \\
\# MRI slices & 1871 \\
tallest height & 115.13 cm \\
shortest height & 110.31 cm \\
median radius &  4.72 cm $\pm$ 0.01 cm \\
median circumference & 29.64 cm $\pm$ 0.08 cm 
\end{tabular}
\caption{Sweet cherry specimen statistics. Uncertainty is the standard error of the median.}
\label{tab:dimensions}
\end{table}

\subsection{Image Processing}
\label{ssec:ImageProcessingBkgd}
MRI generates a series of \verb|dicom| files corresponding to slices of the object being scanned.  Each slice was taken parallel to the the ground so that each is an image of rings of the tree at a fixed height. For ease of notation, we rescale the intensity values over all pixel arrays in a scan so that they span $[0,1]$. 


Mathematically, a $m \times n$ pixel slice (or ``slice'') $Z$ is a function $$Z : \{0,1,\dots,m-1\} \times \{0,1,\dots,n-1\} \to [0,1].$$  This function maps pixel positions to intensity value. For example, if the the pixel in the upper-left hand corner of the slice $Z$ has intensity value $0.5$, then we write $Z(0,0) = 0.5$. Notice that the notion of a pixel slice is equivalent to a 2D image, making it amenable to methods from the field of image processing.

\subsubsection{Thresholding}
\textit{Thresholding} is a fundamental image processing technique which generates a binary image intensity threshold. Given a pixel slice $Z$, select a threshold $\alpha \in [0,1]$. Then define a thresholded slice $Z^{\alpha}: \{0,1,\dots,m-1\} \times \{0,1,\dots,n-1\} \to \{0,1\}$ by 
$$Z^{\alpha}(i,j) = \begin{cases}0&\text{if } Z(i,j)<\alpha \\ 1&\text{if } Z(i,j)\geq\alpha \end{cases}.$$
Thresholding can be used to segment an image or reveal features of interest by eliminating pixels whose value is below $\alpha$. There are many ways to select a threshold. We use Otsu's method, which assumes an underlying distribution of intensity values from two classes and iterates to minimize intra-class variance \cite{otsu_threshold_1979}.

\subsubsection{Edge Detection}
Given an input image, edge detection algorithms generate a new image marking probable edges. The \textit{Sobel edge detection algorithm} works by convolving an image with the Sobel operator to approximate the image gradient, which highlights areas of abrupt change in intensity \cite{gupta_sobel_2013}. Previously the Sobel operator has also been used for tree ring identification in images of cross-sections \cite{henke_semiautomatic_2014}. We apply the Sobel algorithm to estimate the radii of slices in \cref{sec:radius}.

\subsubsection{Graphs and Breadth-First Search}
In order to traverse pixel locations in a pixel slice, we also think of each slice as a mathematical graph. A graph is a mathematical construct consisting of a set of objects and connections between them. The objects are called \textit{edges} and the connections are called \textit{vertices}. A graph $G$ is usually denoted by $G = (V,E)$, where $V$ is the set of vertices and $E$ is the set of edges.

To represent a slice as a graph, let $V$ be all coordinate pair inputs (pixels). An edge exists between two coordinate pairs if exactly one entry in the pairs differs by exactly one. In other words, there exists an edge from a pixel to its neighbors in the up, down, left and right directions. More formally, let $Z$ be an $m \times n$ pixel slice and define a set of directions $D = \{(0,1),(1,0),(0,-1),(-1,0)\}$. Then define the \textit{slice graph} $G= (V,E)$ by
\begin{align*}V &:= \{(i,j) \mid i < m-1; j< n-1\},\\
E &:= \{(i+d_i,j+d_j) \mid (i,j) \in V;\; (d_i,d_j)\in D; \; i+d_i \leq m; \; j+d_j \leq n \}.
\end{align*}
The restrictions $i+d_i \leq m$ and $j+d_j \leq n$ on elements of the vertex set $V$ ensure that there actually exists a pixel neighboring $(i,j)$ in the direction given by $(d_i,d_j)$. 

Using a slice graph to represent a pixel slice allows us to traverse the image and look at pixels in a certain order. In particular, in part of our procedure we use \textit{breadth-first search} (BFS), which we will summarize here. For a more detailed explanation of BFS in image processing, see \cite{silvela_breadth-first_2001}.

Search algorithms for graphs differ by the order in which we look at pixels. BFS is a search algorithm for graphs which looks at nearby neighbors before considering neighbors that are farther away (``breadth-first'' is in opposition to ``depth-first'').  In practice, on a slice graph, this means that if we are seeking a pixel that meets some condition nearby a seed $(x,y)$, we first look at the the intensity of its neighbors. After considering each neighbor of $(x,y)$ we look at the neighbors of the neighbors of $(x,y)$. Depending on the application, this process continues until we find a pixel the meets the condition or we find the largest region containing $(x,y)$ that meets the condition.  
This technique will be used in Sec.~\ref{ssec:ImageProcessingBkgd} to determine the centroid of the pith in each slice.

\subsection{Mathematical Phyllotaxy}
\label{ssec:PhyllotaxyBkgd}
The goal of mathematical phyllotaxy is to describe emergent spiral and other patterns of lateral organs. For the cylindrical representation, this includes viewing phyllotactic patterns as helices on a cylinder or sets of parallel lines.  The idealized geometric model, describing the placement of the primordia and parastichies is described by the Fundamental Theorem of Phyllotaxy \cite{jean_mathematical_1987,swinton_fundamental_2012}, from which Jean's pattern determination table can be used to estimate phyllotactic parameters; e.g., using an allometry-based model \cite{jean_allometric_1983}. Fibonacci numbers associated with parastichies can be seen as fixed points in dynamical systems derived from optimal packing assumptions \cite{atela_dynamical_2003}. For a full review of links between mathematical and molecular phyllotaxy, see \cite{pennybacker_phyllotaxis:_2015}.

Besides their ubiquity in developing plant structures, phyllotactic features appear in other natural phenomena such as capillary structures during evaporation \cite{chen_control_2017} and the cylindrical phyllotaxy of carbon nanontubes \cite{fan_principles_1995}. For a discussion of the universality of Fibonacci patterns in nature, see \cite{shipman_how_2011}. In what follows we define the several phyllotatic parameters used to summarize the arrangement of plant organs.


\subsubsection{Parastichy}
Adopting a convention from \cite{jean_phyllotaxis_1994}, we call the plant organs that comprise phyllotactic patterns \textit{primordia}. In the case of our work, the primordia are epicormic buds near the epidermis connected by radial, vascular traces to the pith. 

A \textit{parastichy} is a set of primordia that form a spiral arm. All parastichies that run in the same direction form a \textit{family} and a parastichy in a family of $n$ spirals is an \textit{$n$-parastichy}. \cref{fig:phyllo} shows families of 5- and 8-parastichies in synthetic data. The 1-parastichy is called the \textit{genetic spiral} and contains all primordia. A \textit{contact parastichy} is one that is derived from the shape of the primordia. For example, each hexagonal primordia on the outside surface of a pineapple suggests three contact parastichies.  


A \textit{parastichy pair} $(m, n)$ is formed by $m$ parastichies in one direction and $n$ parastichies in the other.   Parastichy pairs are often consecutive numbers in a Fibonacci-type sequence. Define a sequence $(F_i)_{i\in \mathbb{N}}$ by $F_1 = 1$, $F_2=t$ and $F_i = sF_{i-1}+sF_{i-2}$, where $t\geq 2$ and $s\geq 1$. In \textit{normal phyllotaxy} parastichy numbers come from the sequence $(F_n)$  and Fibonacci numbers arise in the case that $t=2$ and $s=1$ \cite{jean_phyllotaxis_1994}. We restrict ourselves to the Fibonacci sequence $(1,2,3,5,8,\dots)$.

 A \textit{visibly opposed} parastichy pair intersects only at primordia. There are many of these pairs in a phenomenon known as ``rising phyllotaxy''.  In order to choose a definite parastichy pair, we are interested in the \textit{conspicuous parastichy pair}, which is a visibly opposed parastichy pair such that the angle of intersection between the two families is closest to \ang{90} \cite{jean_phyllotaxis_1994}. This pair is not necessarily unique, but it is the most ``conscipuous'' in the sense that the corresponding families of parastichies are the most noticeable.

\subsubsection{Parameters}
 Since the cherry tree trunk is cylindrical, we limit our discussion of phyllotactic parameters to cylindrical phyllotaxy.   In the cylindrical representation the coordinates of the primordia on the surface of the cylinder are given by
$$(z_0,\theta_0),(z_2,\theta_2),\dots,(z_{q-2},\theta_{q-1}),$$
where $\theta_i \in [0,2\pi)$ for  $i=0,1,\dots,m-1$ (this notation implies an arbitrary position for the ray $\theta=0$). The \textit{divergence angle} $d$ is defined as the average difference between successive angle measurements, i.e.
$$d :=\frac{1}{q-2} \sum_{i=1}^{q-2} \left[(\theta_{i+1} - \theta_i) \text{ mod } 2\pi\right].$$ For clarity we also define the \textit{divergence fraction} as $d^* := d/2\pi$, which is the fraction of the angular breadth of the arc made by the divergence angle (in some literature this is the definition of the divergence angle \cite{jean_phyllotaxis_1994}).


Another useful parameter is the vertical distance between successive primordia, or the \textit{rise}, defined by the internode distance,
$r_i := z_{i} - z_{i-1}$.
The conspicuous parastichy pairs are a function of both  the divergence angle and the rise. Hence, both parameters are necessary to fully to describe the phyllotaxy of the system.

Using the divergence angle $d$, spiral nodes can be generated by $(x_n,y_n) = (n\cos{nd}, n\sin{nd})$ and cylindrical lattice points can be generated by $(z_n,\theta_n) = (n, (dn)/(2\pi) \mod{2\pi})$. Examples of both centric and cylindrical phyllotaxy are shown in \cref{fig:phyllo}. 



\subsubsection{Phyllotactic Fractions}\label{sec:frac}
In contrast to paristichy pairs, a more traditional way to describe a cylindrical phyllotactic system is to use a phyllotactic fraction to approximate the angular differences between nodes \cite{allard_aspects_1942}. A \textit{phyllotactic fraction} $A/B $ (sometimes called ``the phyllotaxis'' of a species) is determined by the number of turns of a spiral, $A$, of successive leaves to reach the $B$th node directly above the starting node. 
In the fraction, $A$ and $B$ are always every other number in a Fibonacci-type sequence from normal phyllotaxy. For example, cherry’s phyllotactic fraction, derived from the Fibonacci sequence, is $2/5$; this means that a node will have a node form above it after approximately two spirals of five successive nodes. A cherry stem will then have five vertical ranks of nodes called \textit{orthostichies} \cite{okabe_biophysical_2015}. The phyllotactic fraction is meant to approximate the divergence angle. For example, the phyllotactic fraction 2/5 implies a divergence angle of $d = (2/5)\ang{360} = \ang{144}$. 

\subsubsection{Fundamental Theorem of Phyllotaxy}
Using the divergence fraction, we can identify parastichies in an idealized model describing the pattern of primordia.

First, \cref{thm:bravis} helps identify which primordia belong to which parastichies. After sorting the primordia by their radial angle and then labelling the primordia by their position, we call two primordia \textit{adjacent} if they are next to each other in this list.
\begin{Theorem}[The Bravais-Bravais Theorem, 1837 \cite{jean_phyllotaxis_1994}]\label{thm:bravis}
On an $n$-parastichy of a phyllotactic spiral pattern, the numbers of two adjacent primordia differ by $n$.
\end{Theorem}
It follows from \cref{thm:bravis}, for example, that the parastichies in the 2-parastichy family will have labels $\{0,2,4,\dots,\}$ and $\{1,3,5,\dots,\}$.  To identify which parastichies are visibly opposed is characterized by \cref{thm:fund}.
For ease of notation let $\text{nint}(x)$ denote the nearest integer to the real number $x$.

\begin{Theorem}[Fundamental Theorem of Phyllotaxy, 1994 \cite{jean_phyllotaxis_1994}; Revised 2012 \cite{swinton_fundamental_2012}] \label{thm:fund}
The following are equivalent:
\begin{enumerate}
\item The parastichy pair $(m,n)$ is visible and opposed, where $m \leq n$ and $\Delta = \textup{nint}(nd^*)m - \textup{nint}(md^*)n$.
\item \begin{enumerate}
    \item $m=1$, $n=1$, $d^* = 1/2$ and $\Delta = 1$, or
    \item $m=1$, $n > 1$, $\Delta = 1$, $d^* \in (1/2n, 1/n)$, or
    \item $m=1$, $n>1$, $\Delta = -1$, $d^* \in (1 - 1/2n, 1 - 1/n)$, or
    \item $1 < m < n$, $d^* \in (u/m, v/n)$, $\Delta = \pm 1$, where $u,v$ are the unique integers $0 \leq v < n$, and $0 \leq u < m$ such that $mv - nu = \Delta$.
\end{enumerate}
\end{enumerate}
\end{Theorem}

Notice that bounding the divergence angle in a particular way is necessary and sufficient to determine if a parastichy pair $(m,n)$ is visible and opposed. The following theorem restates the \cref{thm:fund} in terms of the Fibonacci sequence.
\begin{Theorem}
[Adler's Theorem, 1974  \cite{adler_model_1974}]
\label{thm:adler}
 Let $(F_k)_{k\in \mathbb{N}}$ denote the Fibonacci sequence and define an interval $$I_k = \begin{cases}
 [F_{k-2}/F_k, F_{k-1}/F_{k+1}], &$k$ \text{ odd} \\
  [F_{k-1}/F_{k+1}, F_{k-2}/F_{k}], &$k$ \text{ even} 
 \end{cases}.$$ The parastichy pair $(F_k,F_{k+1})$ is visible and opposed if and only if $d^* \in I_k$.
\end{Theorem}
Note that $\bigcap_{k=1}^{\infty} I_k = \{\phi^{-2}\}$, where $\phi$ is the golden ratio. Hence, if $(F_k,F_{k+1})$ is visble and opposed for all $k$, then $d^* = \phi^{-2} \approx  0.381$. The golden angle is defined as  $2\pi\phi^{-2} \approx \ang{137.508}$.

\section{Results}
\label{sec:Methods}
Using image processing techniques and mathematical phyllotaxy, we propose a method to algorithmically determine the conspicuous parastichy pair found in cylindrical phyllotaxy. This method could be the basis of an automatic method to retrieve the locations of primordia in 3D scans of plants with cylindrical phyllotaxy, given that the primordia are revealed by the distribution of intensity values. 
Again let $\text{nint}(x)$ denote the nearest integer to the real number $x$.

\subsection{Pre-Processing}
The cherry tree scan is comprised of 1871 dicom files, each of which contains a $250 \times 250$ pixel slice.  Each slice represents a thickness of 0.625 mm and each pixel a 0.351562 mm $\times$ 0.351562 mm square. Hence, each voxel cube in the image has a volume of about 0.077 mm. Using the \verb|pydicom| Python package \cite{dicom}, we converted dicom files to \verb|numpy| matrices \cite{numpy}.


\subsection{Pith Finding Algorithm}
\label{sec:pith}
Since the tree trunk is not a perfect cylinder, the location of the tree's pith in each slice is different. 
To be able to orient the object in a coordinate system, we first identify the location of the pith centroid in each slice which we will use later to convert each slice to polar coordinates. 

To keep the description of procedure as general as possible, suppose we have a scan with $N$ slices denoted by $Z_1,Z_2,\dots,Z_N$. 
Given the center of the pith in $Z_1$, we iteratively locate the center of the pith in slices $Z_2,\dots,Z_N$ using a search algorithm. 
First define a range of intensity values $C \subset [0,1]$ which represents the intensity values found in the pith in the whole scan, since the pith is a \textit{region} (collection of coordinate pairs) and just one coordinate pairs. In our image, we chose $C = [0,0.6]$. Assuming  there exists a boundary around the target pith region in each slice whose intensity values are not in $C$, we can use $C$ to search for the pith region. 

We require user input to determine a starting location for the pith in the first slice, $Z_1$. 
As this will likely not be exactly the centroid of that pith, we will call this initialized input $(x_{0},y_{0})$.
Let $(x_{i-1}, y_{i-1})$ be the center of the pith in slice $Z_{i-1}$. 
Using BFS in slice $Z_i$, we spread outward from $(x_{i-1}, y_{i-1})$ until we find a coordinate pair $(\hat{x}, \hat{y})$ such that $Z_i(\hat{x}, \hat{y}) \in C$. 
Note that it is possible for $(\hat{x}, \hat{y})$ to be the starting coordinate $(x_{i-1}, y_{i-1})$ if $Z_i(x_{i-1}, y_{i-1}) \in C$.
This puts us inside the pith region of slice $Z_i$. 
We find the whole pith region by using another BFS to find the entire connected region containing $(\hat{x}, \hat{y})$, such that all each coordinate pairs in the region have intensity values in $C$. 
Finally, the centroid of this region is marked as the center of the pith for slice $Z_i$. 
The full algorithm is outlined in \cref{algo:pith}.


\begin{algorithm}[H]
 \KwData{Slices $Z_1,Z_2,\dots,Z_N$, initial center $(x_0,y_0)$, intensity range $C \subset [0,1]$}
 \KwResult{ Center coordinates $(x_i,y_i)|_{i=1}^{N} $}
 \For{i=1,\dots,N}{
Using BFS, choose 
$(\hat{x},\hat{y}) \in 
\text{argmin} \{\|(x,y) - (x_{i-1},y_{i-1})\|
\mid Z_i(x,y) \in C\}$ \\
Using BFS, find $R$, the largest connected region of $Z_i$ such that $(\hat{x},\hat{y}) \in R$ and $Z_i(x,y) \in C$ for all $(x,y) \in R$ \\
Set $n = |R|$ to be the number of pixels in $R$ \\
Set $x_i :=\text{nint}\left(\frac{1}{n} \sum_{(x,y)\in R}x \right)$ and $y_i := \text{nint}\left( \frac{1}{n} \sum_{(x,y)\in R}y \right)$
 }
 \caption{Algorithm for finding pith centroid locations}
 \label{algo:pith}
\end{algorithm}

\subsection{Radius Estimation}
\label{sec:radius}
To estimate the average radius of the cherry tree we identify the edge of the tree in each slice and then compute the distance to the pith location. This process is outlined in \cref{fig:sobel}.  
Specifically we apply the Sobel edge detection algorithm to identify probable edges. 
Since the edge detection algorithm also identifies rings and traces in the interior of the tree, we threshold the edges and keep only values above the 99th percentile, leaving values on the boundary of the tree. 
Then we compute the median distance from these points to the pith centroid to estimate the radius of a slice. 
Let $\rho_i$ denote the estimated radius of slice $Z_i$. 
We estimate the overall radius of the tree as the median of $\rho_i$ over all $i$.

\subsection{Polar Conversion}
Using the coordinate pairs for pith locations produced by \cref{algo:pith} and the estimated radii, we convert each slice to polar coordinates. 
Then the radial, wedged-shaped traces in the slices become vertical blocks and are easier to identify.

Specfically, for each slice $Z_i$, define a \textit{polar slice} $P_i$ by
$$P_i(r,\theta) := Z_i(\text{nint}( x_i + r\cos\theta), \; \text{nint}(y_i + r\sin\theta))$$ for $\theta \in \{0,\alpha,2\alpha,\dots,2\pi-\alpha\},  r \in \{0,s,2s,\dots,\rho_i-s\}$, where $\alpha,s$ are small positive numbers.  
Selected pixel slices and polar slices are shown in \cref{fig:slices}.


\subsection{Boundary Image}
Next we construct a radial summary of the polar slices that we call the \textit{boundary image}.  Notice that the traces in \cref{fig:slices}. have higher intensity values than the other parts of the tree. This means the intensity values in a column will tend to be higher if a trace is present. Hence, a summary statistic of the intensity values in each column may capture whether or not a trace exists there.

For each polar image $P_i$, define an intensity  range $T_i \subset [0,1]$ that denotes the foreground of the image. Define the boundary image $\mathcal{B}:\{1,\dots,N\} \times \{1,\dots,\floor{\frac{2\pi}{\alpha}}\} \to [0,1]$ by  $$\mathcal{B}(i, \theta) := \text{mean}_r\{P_i(r,\theta) \mid P_i(r,\theta) \in T_i \}.$$
Hence, the mean pixel intensity along columns in polar slices become rows in the boundary image (the intensity range $T_i$ restricts us to consider only pixels that are a part of the tree and not pixels that are in the background). Note that other summary statistics such as the median or the $p^{th}$ percentile may be used in place of the mean. 


The boundary image is shown in \cref{fig:features}(A). The boundary rays $\theta = 0$ and $\theta = 2\pi$ glue together so that the boundary image is the boundary of a cylinder representing the cherry tree. Notice the phyllotactic patterns that emerge in the lattice structure of the image.


\subsection{Blob Detection}
Each blob in the boundary image corresponds to an epicormic trace in the cherry tree. To identify the centroid of each blob, we threshold the image using Otsu's method. Then we compute the centroid of each region to find the trace locations, shown in \cref{fig:features}(B). Notice that one blob (colored in red) was too small to be recognized. The centroid of this blob was found by using a less restrictive threshold and was added to the final dataset.

\subsection{Phyllotaxy Parameter Determination}
Using the centroids of the blobs in \cref{fig:features}(B), we estimate phyllotactic parameters in the context of both the Fundamental Theorem of Phyllotaxy (\cref{thm:fund}) and phyllotactic fractions. Start by sorting and labeling the 36 nodes by their $z$ location so that node 0 is near the bottom of the trunk and node 35 is near the top. The divergence angle, estimated by the mean change in angle between successive primordia is $d =$ \ang{142.928} $\pm$ \ang{1.181}, which gives a divergence fraction of $d^* =$ 0.397 $\pm$ 0.003. To see the distribution of local divergence angles and rise, see \cref{fig:params}.  
Assume that $d^*$ is the actual divergence fraction of the system. Since $d^* \in [3/8, 2/5]$, it follows from  Adler's Theorem (\cref{thm:adler}) that $(1,2), (2,3), (3,5), (5,8)$ are all visible and opposed parastichy pairs. Also, 1, 2, 3, 5 and 8 are parastichy numbers in the sense that are part of visibly opposed parastichy pairs for the specimen.  However, the standard error on $d^*$ suggests that the true mean could be greater than $2/5$, which would exclude the 8-parastichies from consideration.  \cref{fig:3D} shows all parastichies in 3D and on the Bravais lattice.

Since the second number in a phyllotactic fractions corresponds to a parastichy number, 1/2, 1/3, 2/5 and 3/8 are all phyllotactic fractions of the system (see \cite[§2.2.2]{jean_phyllotaxis_1994}).  Traditionally, the phyllotactic fraction $A/B$, chosen to represent the phyllotaxy of a cylindrical system, is derived from the following observation: take two nodes whose labels differ by $A$ and note that they are approximately above each other (having the same $\theta$ values) by going exactly $B$ times around the genetic spiral. In this sense, the ``phyllotaxy of the system'' for the cherry specimen can be described as 2/5.  We propose a more precise way of determining a single phyllotactic fraction: take the Fibonacci fraction closest to the divergence fraction $d^*$. 
In other words, for phyllotaxy based on the Fibonacci sequence, let $F_{I}/F_{I+2}$ be the \textit{conspicuous phyllotactic fraction}, where $I = \text{arginf}_i\,|(F_i/F_{i+2}) - d^*|$.  In our case, $I = 3$, $F_3 = 2$ and $F_5=5$ since 2/5 is closest to $d^* = 0.397$. 

To determine the conspicuous parastichy pair, we calculate angles between parastichies. 
Using the Bravais-Bravais Theorem (\cref{thm:bravis}), draw $F_{i}$-parastichies and $F_{i+1}$-parastichies using linear interpolation and measure the angles at each intersection.  
The pair of $F_i$ and $F_{i+1}$ for which this angle is closest to $\ang{90}$ is the conspicuous parastichy pair (all parastichy families are shown in \cref{fig:3D}). Using the law of cosines, we compute the average intersection angles between families of parastichy pairs and convert each to the ``small angle'': \ang{77.91} for (2,3); \ang{35.21} for (3,5); and \ang{14.75} for (5,8). Thus, by definition, the parastichy conspicuous parastichy pair is $(F_2,F_3) = (2,3)$. The 2-family and 3-family of parastichies are shown in the Bravais lattice in \cref{fig:features}(C). A summary of phyllotactic parameters found for this sample are shown in  \cref{tab:params}.


\begin{table}[!ht]
\centering
\begin{tabular}{lllll}
\textbf{Parameter} & \textbf{Value} \\
number of primordia &  36 \\
handedness & counterclockwise\\
divergence angle ($d$) & \ang{142.928} $\pm$ \ang{1.181} (2.495 rad $\pm$ 0.021 rad) \\
divergence fraction ($d^*$) & 0.397 $\pm$ 0.003  \\
average rise ($r$) &  2.896 \text{cm} $\pm$ 0.182 cm \\
visible \& opposed parastichy pairs & (1,2), (2,3), (3,5), (5,8)\\
parastichy numbers & 1, 2, 3, 5, 8\\
conspicuous parastichy pair &  (2,3) \\
conspicuous phyllotactic fraction & $F_3/F_5 = 2/5$
\end{tabular}
\caption{Phyllotactic parameters with standard error. The ``handedness'' of the system is the direction of the genetic spiral going from the bottom of the sample. The families of parastichies alternate between clockwise and counterclockwise direction. }
\label{tab:params}
\end{table}

\section{Discussion}
\label{sec:discussion}
The 36 nodes detected in our sample were patterned in a single year. The traces connecting to the epicormic buds traverse eight years of secondary growth. In the orchard, trees are planted as a ``whip'' of a genetically compound tree, that is, a one-year-old nursery tree is comprised of the shoot of the scion genotype (fruiting variety) that grew in the nursery the previous year from a bud that was grafted onto a rootstock genotype. A whip nursery tree is typically 1 to 1.5 m tall. In this experiment, the number (8) of annual growth rings at the top of the scanned trunk section matched the number at the bottom; therefore, each node in the section of the scanned trunk originated in the same year, all on the original whip nursery tree. To our knowledge, this is the first accounting of a single growing season's complement of sweet cherry nodes from origin through eight years of trunk growth that also identified the persistence of epicormic bud traces. That is, the phyllotactic patterning that occurred during juvenile shoot growth remains in a mature tree, and can be detected using a combination of magnetic resonance imaging and image processing approaches.
    
\subsection{Conclusions}
Our results have practical implications for orchard rejuvenation and directed approaches to influence tree architecture. As orchards age, yield and fruit quality can begin to decline if fruiting sites become shaded and/or portions of the tree become infected by diseases. Epicormic buds serve as a ``bank'' for new branches to sprout and rejuvenate orchards. Magnetic resonance imaging provides a clear picture of how full that bank was after eight years. This could facilitate study of how persistence of epicormic buds may be affected by cultivar, vigor, harsh winters, drought, disease, and spring freezes. Determination of phyllotaxis provides growers with the potential to identify where an epicormic bud is located so that orchard training measures may be attempted to force a branch to sprout at that location. This could help growers fill ``holes'' in tree canopies to increase fruiting potential and lengthen the life of the orchard. The study of epicormic structures, which are hidden within secondary growth, using tomographic methods also opens the possibility of studying the genetic and environmental basis of such structures.

Our results also provide an empirical way to measure phyllotactic parameters. Much work in phyllotaxy has focused on generative models. But the image processing techniques presented here provide a method to isolate nodes and place them in the context of shoot growth using anatomical features. This is even possible if the features are difficult to discern by eye, or embedded internally within extensive secondary growth. With larger sets of node locations, Fourier Methods could further automate the process of finding parastichy numbers \cite{negishi_determining_2017, liew_localized_2011}. Isolating features in mature specimens can lead to insights regarding the developmental history of a plant, which is crucial to manipulate plant forms in a directed way and mechanistically understand the origins of morphology. Automated methodology to model imaging features—--from MRI or otherwise—--into a developmental context is an important first step towards an empirical mathematical framework for measuring plant morphology.


\begin{backmatter}

\section*{Author's contributions}
ME, JL, GL, DHC, and EM designed the project and contributed to writing and editing the manuscript. JL and GL collected plant materials and performed the MRI scan. ME, DHC, and EM analyzed data. ME and EM developed new analysis tools and performed mathematical analyses.


\section*{Competing interests}
The authors declare that they have no competing interests.

\section*{Availability of data and materials}
Codes are available on Github (https://github.com/eithun/cherry-phyllotaxy), and raw data are available on the figshare repository (https://doi.org/10.6084/m9.figshare.7409843).

\section*{Funding}
This project was supported by the USDA National Institute of Food and Agriculture, and by Michigan State University AgBioResearch.  The work of EM was supported in part by NSF grants DMS-1800446 and CMMI-1800466.


\bibliographystyle{bmc-mathphys} 
\bibliography{bmc_article}      




\section*{Figures}

\begin{figure}[!ht]
\centering
\includegraphics[height=1.75in]{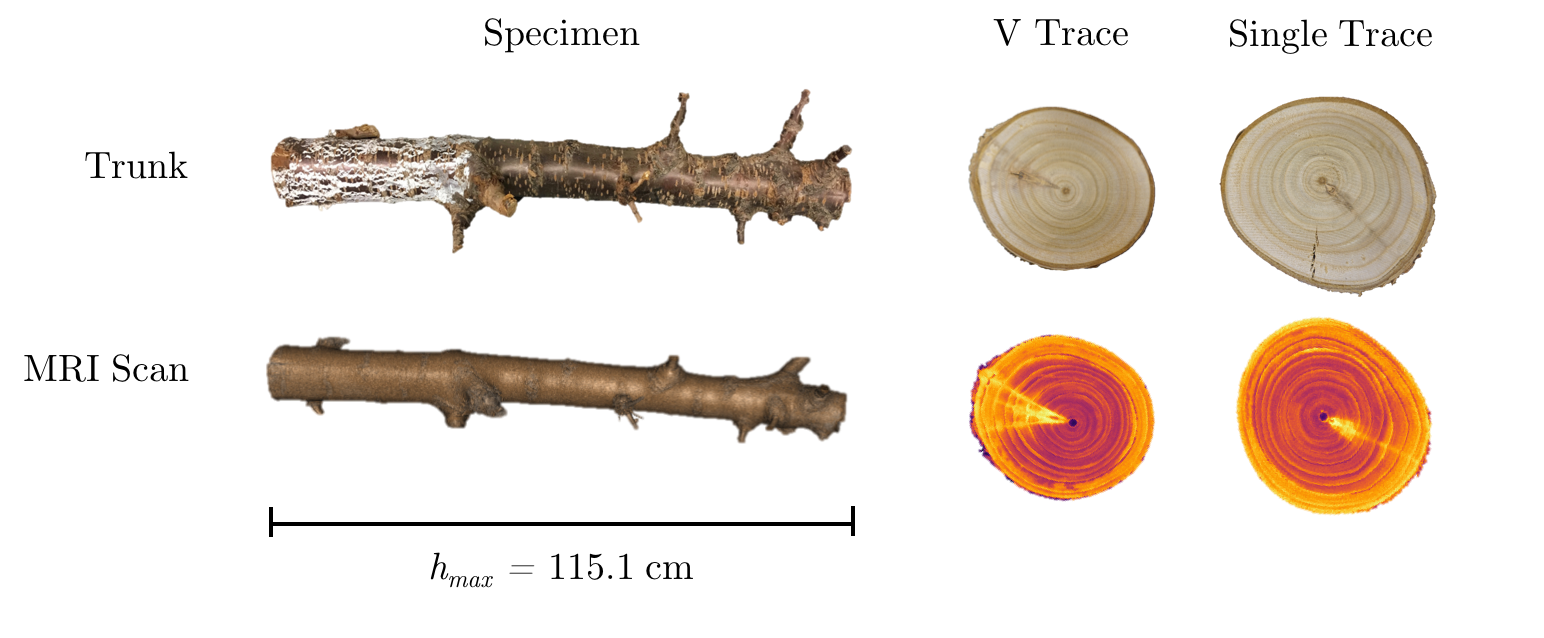}
\label{fig:tree}
  \caption{\csentence{Cherry tree trunk and 3D reconstruction.}
  MRI produces a voxel-based image of the specimen. Using 3DSlicer, an open-source software tool, we created a 3D reconstruction. Two types of traces (a branching, or "V", trace and a single trace) are shown as physical cross-sections made by cutting the trunk and as slices from the scan.}
\end{figure}

\begin{figure}[!ht]
\centering
\includegraphics[height=1.7in]{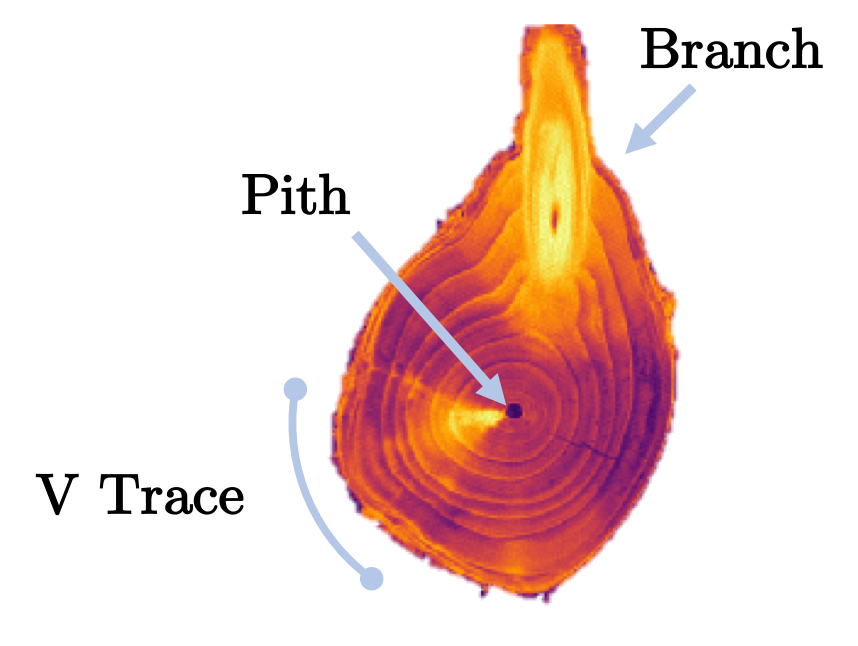}
\caption{\csentence{Anatomy of sweet cherry wood.} A labeled slice from the MRI scan of sweet cherry, colored by intensity value. The pith is the hollow center of the trunk. Both the branch and the epicormic trace are high intensity regions.}
\label{fig:parts}
\end{figure}

\begin{figure}[!ht]
\centering
\includegraphics[height=2.25in]{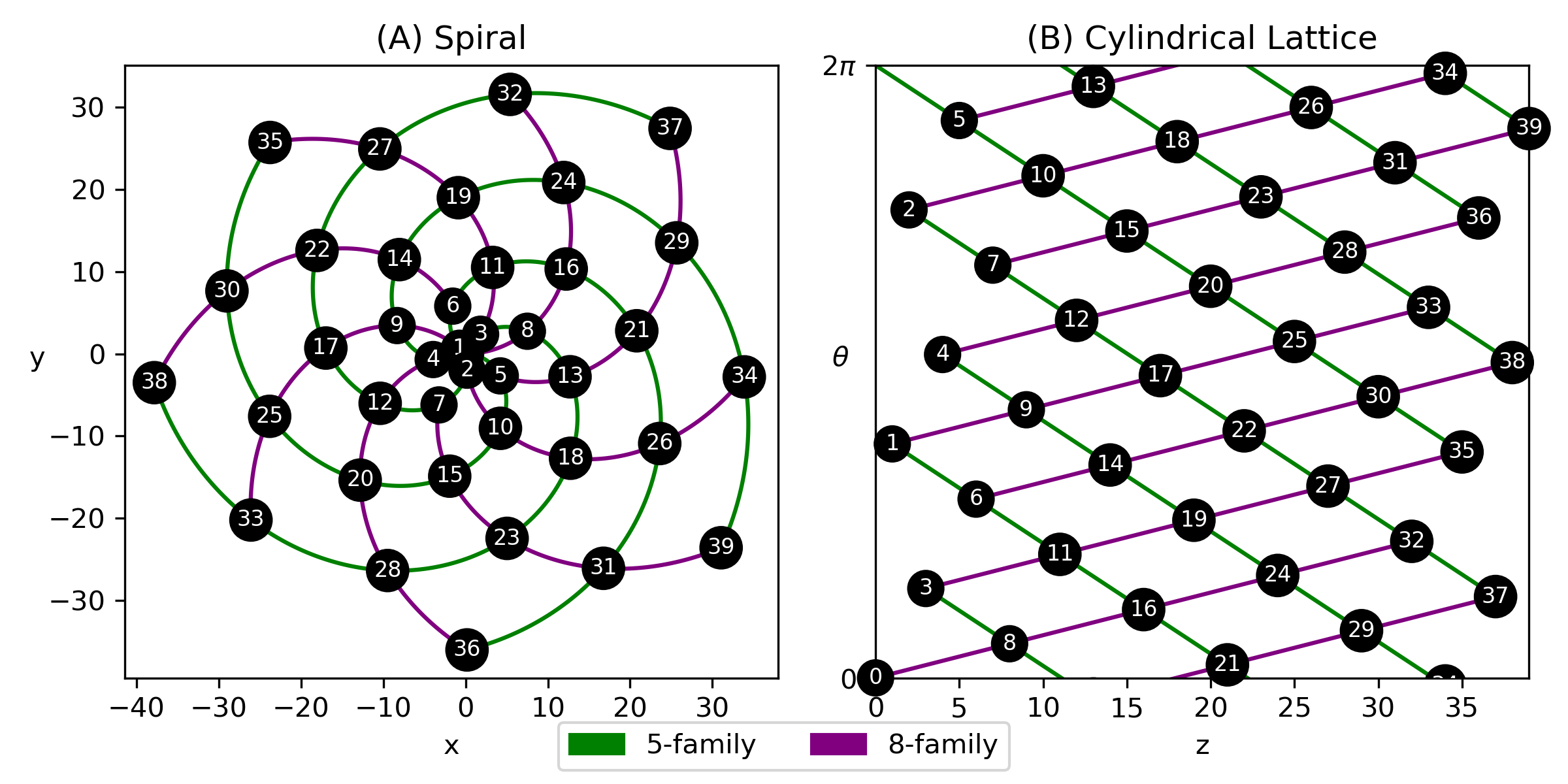}
\caption{\csentence{Two views of parastichy spirals from artificial data.} A set of 40 nodes, generated by the golden angle and constant rise ($r=1$). In the centric form, parastichies are spiral arms emanate from the origin and in the cylindrical form, parastichies are helices in a cylindrical lattice, called the Bravais lattice. In the Bravais lattice, the $\theta = 0$ and $\theta=2\pi$ rays are the same, allowing the parastichies to ``wrap around'' the circumference of the plant.}
\label{fig:phyllo}
\end{figure}

\begin{figure}[!ht]
\centering
\includegraphics[width=0.9\textwidth]{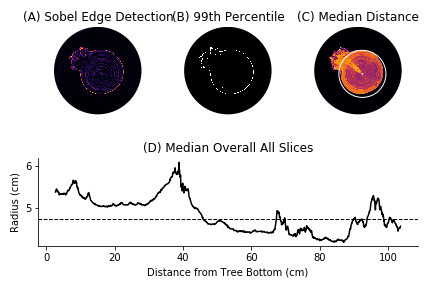}
\caption{\csentence{Algorithm to determine the approximate tree radius in each slice.} (A) To estimate the radius of the trunk in each slice we first apply the Sobel Edge Detection algorithm to reveal areas of abrupt change in pixel intensity. (B) Then we threshold the image with $\alpha$ equal to the 99th percentile of the values given by edge detection. (C) Finally we compute the median distance from the pith location to all of the pixels in the binary image to estimate the radius. (D) To estimate the overall radius for the trunk, we compute the median of all radii for each slice. The dotted line represents the median }
\label{fig:sobel}
\end{figure}

\begin{figure}[!ht]
\centering
\includegraphics[height=3.3in]{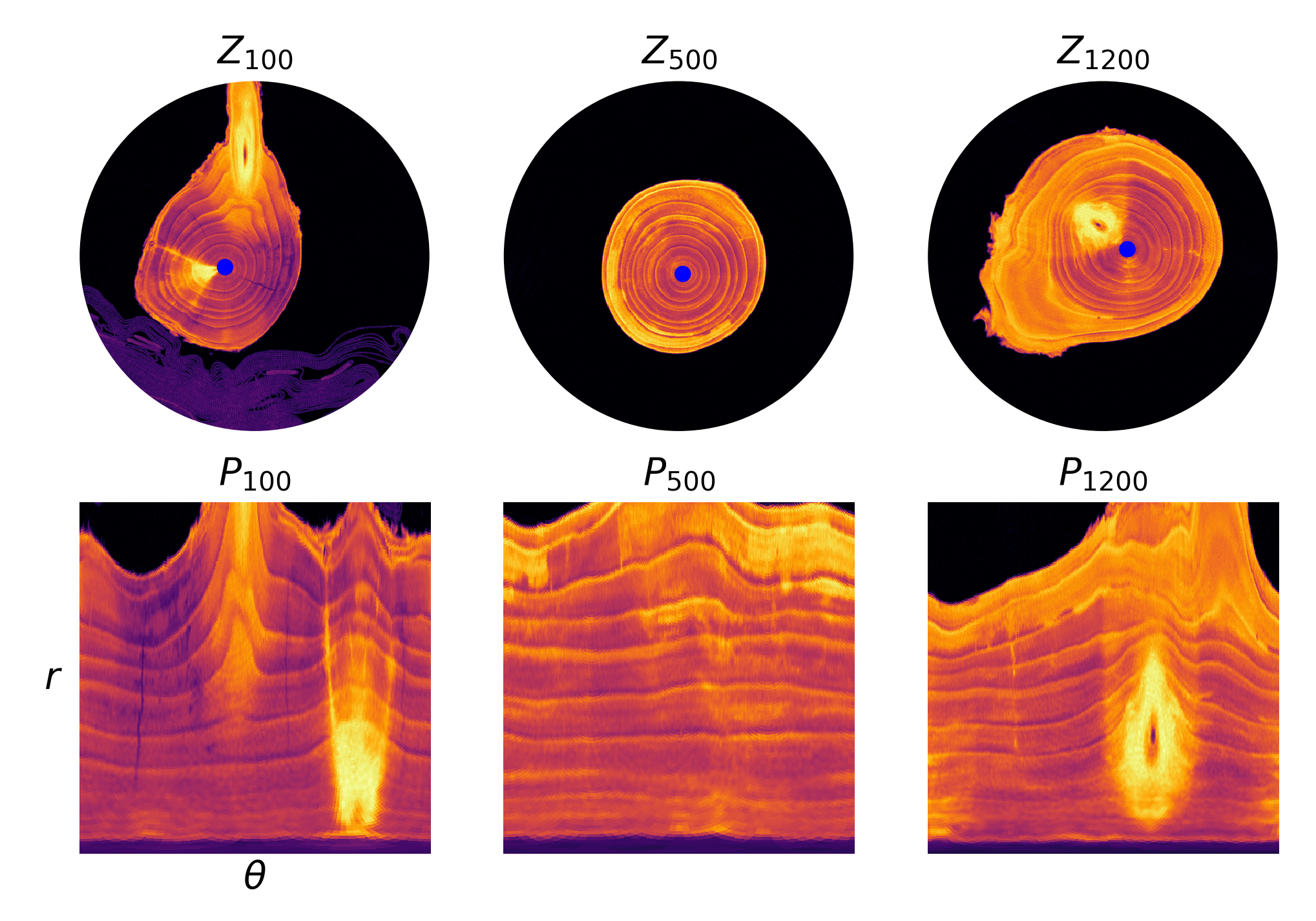}
\caption{\csentence{Polar coordinate conversion for selected slices.} After finding the center and radius of each pixel slice $Z_i$ we convert to polar coordinates, producing polar slices $P_i$. The  radius spans from 0 to the estimated radius $\rho_i$.}
\label{fig:slices}
\end{figure}

\begin{figure}[!ht]
\centering
\includegraphics[height=3in]{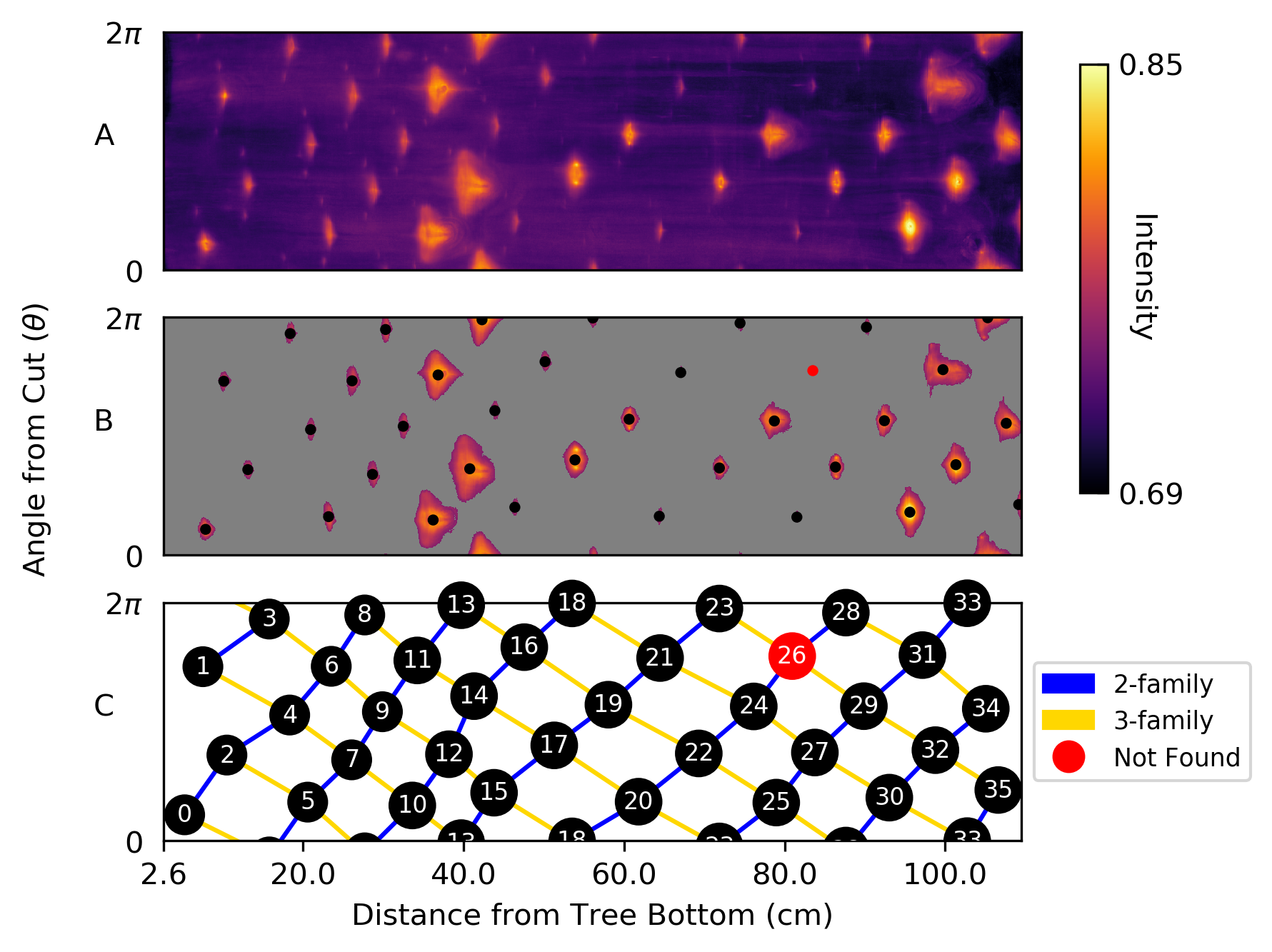}
\caption{\csentence{Isolating phyllotactic features in the boundary image.} (A) The boundary image, created by assembling the means along each radial direction in the polar slices. Rows correspond to polar coordinates and columns slices. (B) The thresholded boundary image, created using Otsu's method. (C) Cylindrical Bravais lattice of nodes, with 2- and 3-parastichies shown. The blob corresponding to node 26 was added by choosing a less restrictive threshold for that region. }
\label{fig:features}
\end{figure}

\begin{figure}[!ht]
\centering
\includegraphics[height=1.5in]{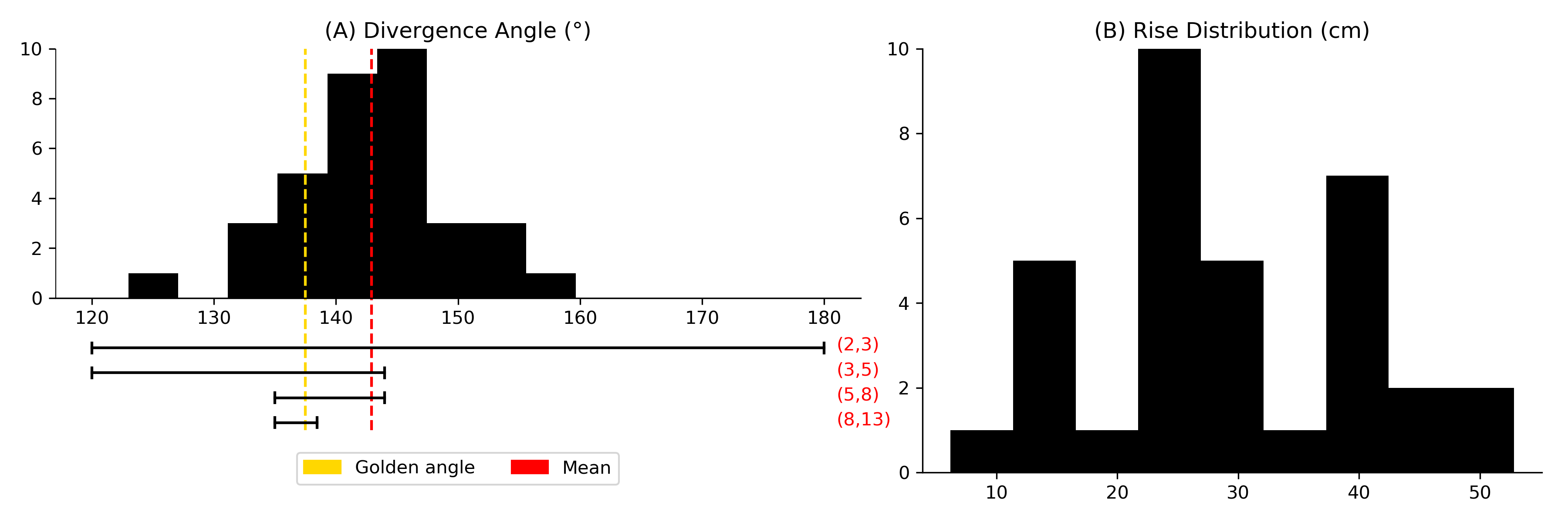}
\caption{\csentence{Distributions of phyllotactic parameters.} (A) The distribution of local divergence angles compared to the golden angle ($\approx \ang{137.5}$) and the mean ($\approx \ang{139}$). The intervals given by Adler's Theorem (\cref{thm:adler}) for visible and opposed parastichy pairs are shown below the distribution. Assuming the mean of the distribution is the true divergence angle, (1,2), (2,3), (3,5) and (5,8), are visible and opposed. The interval [\ang{60}, \ang{180}], corresponding to the pair (1,2), is not shown. (B) The distribution of rise values, showing the range of internode lengths.}
\label{fig:params}
\end{figure}

\begin{figure}[!ht]
\centering
\includegraphics[width=0.9\textwidth]{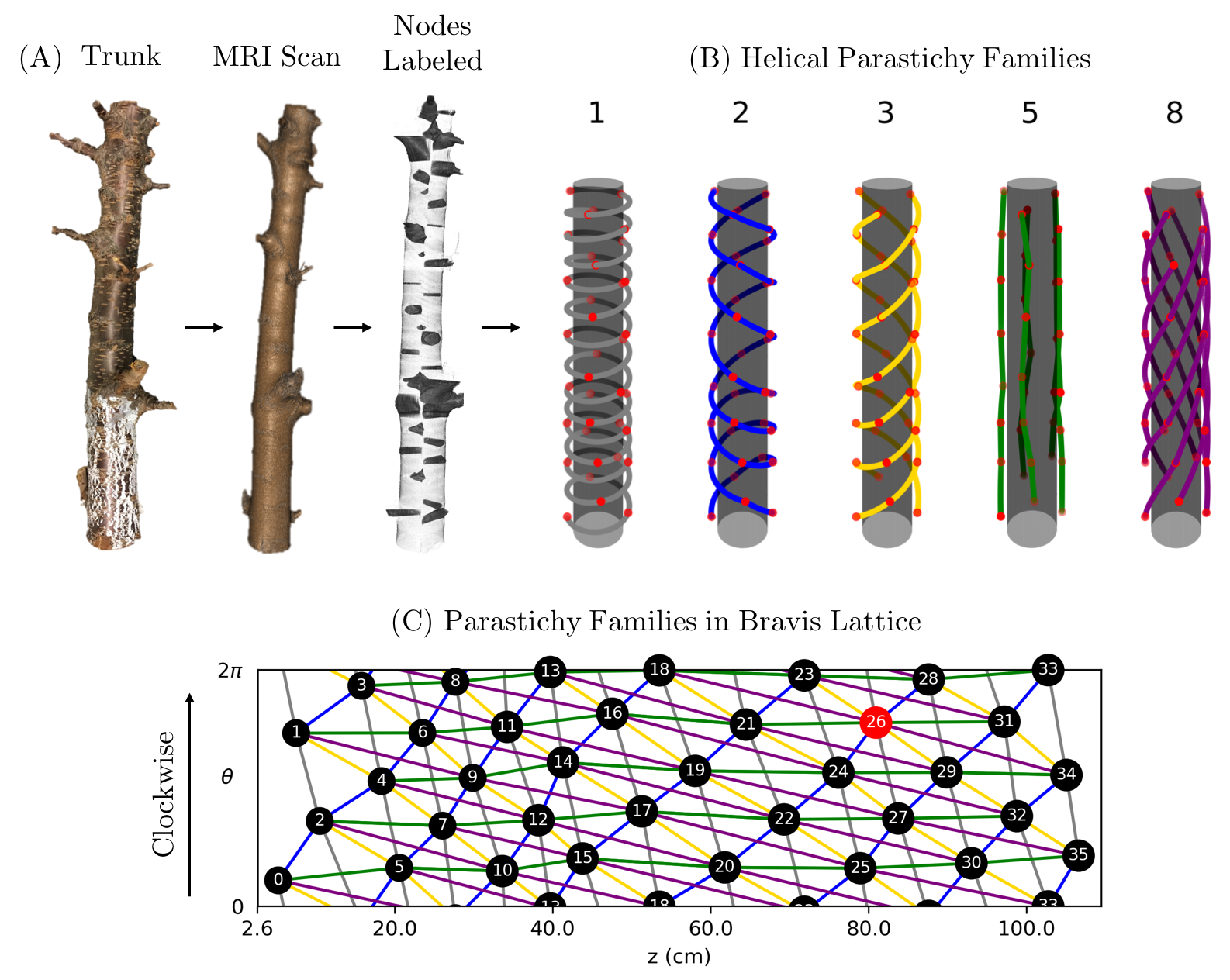}
\caption{\csentence{Visualizing parastichies in 3D.} Families of parastichies using helical interpolation on a cylinder \cite{eberly_fitting_nodate}. (A) The blobs found in \cref{fig:features}(B) map to wedge shaped regions in the MRI scan. (B) A 3D visualization of parastichy families that are a part of visible and opposed  parastichy pairs. (C) Parastichy families shown in the Bravais lattice, with the same color scheme as part (B). }
\label{fig:3D}
\end{figure}







\end{backmatter}
\end{document}